\documentclass[11pt]{article}
\usepackage[english]{babel}

\usepackage[lmargin=1.1in,rmargin=1.1in,bottom=1.3in,top=1.3in,
twoside=False]{geometry}

\usepackage{relsize,xspace}
 \usepackage{xcolor}
 \usepackage{mathtools}
 \usepackage{todonotes}
 \usepackage{comment}
\usepackage{microtype}
\usepackage{amsmath}
\usepackage{amssymb}
\usepackage{amsfonts}
\usepackage{tikz}
\usepackage{bm}
\usepackage{soul}

\usepackage{tikz}

\tikzstyle{vertex}=[circle,inner sep=0.5,minimum size%
=5mm,semithick,fill=black!20, draw=black]
\tikzstyle{S}=[dash pattern=on 2pt off 1pt,blue,xshift=2mm]
\tikzstyle{G}=[black,bend left=10]
\tikzstyle{U}=[red,bend right=10,thick,densely dotted]
\tikzstyle{T}=[green!70!black,densely dashed,very thick]

\usepackage{refcount}
\usepackage{wrapfig}

\usepackage{marginnote}

\definecolor{dblue}{rgb}{0.1,0.2,0.5}
\definecolor{brown}{rgb}{0.6,0.6,0.2}
\usepackage[ocgcolorlinks, linkcolor={dblue}, citecolor={brown}]{hyperref}

\usepackage[amsmath,thmmarks,hyperref]{ntheorem}
\usepackage{cleveref}

\crefformat{page}{#2page~#1#3}%
\Crefformat{page}{#2Page~#1#3}%
\crefformat{equation}{#2(#1)#3}%
\Crefformat{equation}{#2(#1)#3}%
\crefformat{figure}{#2Figure~#1#3}%
\Crefformat{figure}{#2Figure~#1#3}%
\crefformat{section}{#2Section~#1#3}
\Crefformat{section}{#2Section~#1#3}
\crefformat{chapter}{#2Chapter~#1#3}
\Crefformat{chapter}{#2Chapter~#1#3}
\crefformat{chapter*}{#2Chapter~#1#3}
\Crefformat{chapter*}{#2Chapter~#1#3}
\crefformat{part}{#2Part~#1#3}
\Crefformat{part}{#2Part~#1#3}
\crefformat{enumi}{#2(#1)#3}
\Crefformat{enumi}{#2(#1)#3}

\usepackage{enumerate}

\usepackage{latexsym}

\theoremnumbering{arabic}
\theoremstyle{plain}
\theoremsymbol{}
\theorembodyfont{\itshape}
\theoremheaderfont{\normalfont\bfseries}
\theoremseparator{}

\newtheorem{theorem}{Theorem}
\crefformat{theorem}{#2Theorem~#1#3}
\Crefformat{theorem}{#2Theorem~#1#3}

\newtheorem{lemma}[theorem]{Lemma}
\crefformat{lemma}{#2Lemma~#1#3}
\Crefformat{lemma}{#2Lemma~#1#3}
\newtheorem{claim}{Claim}
\crefformat{claim}{#2Claim~#1#3}
\Crefformat{claim}{#2Claim~#1#3}

\def\cqedsymbol{\ifmmode$\lrcorner$\else{\unskip\nobreak\hfil
\penalty50\hskip1em\null\nobreak\hfil$\lrcorner$
\parfillskip=0pt\finalhyphendemerits=0\endgraf}\fi}

\theoremstyle{nonumberplain}
\theoremheaderfont{\itshape}
\theorembodyfont{\normalfont}
\theoremsymbol{\ensuremath{\square}}
\newtheorem{proof}{Proof}

\theoremsymbol{\ensuremath{\lrcorner}}
\newtheorem{clproof}{Proof}

\newcommand{\col}{\mathrm{col}}
\newcommand{\adm}{\mathrm{adm}}

\newcommand{\Sreach}{\mathrm{SReach}}

\newcommand{\CCC}{\mathcal{C}}

\newcommand{\MC}{\textrm{MC}}
\newcommand{\LLL}{\mathcal{L}}
\newcommand{\Pp}{\mathcal{P}}
\newcommand{\FPT}{\ensuremath{\mathrm{FPT}}}
\newcommand{\Oh}{\mathcal{O}}

\def\cqedsymbol{\vbox{\hbox{\hskip1.2ex\vrule height1.2ex width0.6pt}\hrule
    height0.6pt}}

\newcommand{\N}{\mathbb{N}}

\renewcommand{\phi}{\varphi}
\renewcommand{\epsilon}{\varepsilon}
\newcommand{\strA}{\mathfrak{A}}

\newcommand{\FO}{\ensuremath{\mathrm{FO}}}
\newcommand{\FOs}{\mathrm{FO}[\tau_\mathrm{succ}]}
\newcommand{\minor}{\preccurlyeq}
\newcommand{\dist}{\mathrm{dist}}

\renewcommand{\mid}{~:~}

\newcommand{\abs}[1]{\ensuremath{\left\lvert#1\right\rvert}}

\newcommand{\qitem}[1]{\noindent\leavevmode\hangindent1\parindent%
  \noindent\hbox to1\parindent{#1\hss}\ignorespaces}

\author{Jan van den Heuvel\,\thanks{\,Department of Mathematics, London
    School of Economics and Political Science, United
      Kingdom; \texttt{\{j.van-den-heuvel,
      d.quiroz\}@lse.ac.uk}.} \and Stephan
  Kreutzer\,\thanks{\,Technische Universit\"at Berlin, Germany;
    \texttt{\{stephan.kreutzer, roman.rabinovich\}@tu-berlin.de}.}
  \and Micha\l~Pilipczuk\,\thanks{\,Institute of Informatics, University of
    Warsaw, Poland; \texttt{\{michal.pilipczuk, siebertz\}@mimuw.edu.pl}.}
  \and Daniel A.\ Quiroz\,$^\dagger$ \and Roman Rabinovich\,$^\ddagger$
  \and Sebastian Siebertz\,$^\S$}

\title{Model-Checking for Successor-Invariant First-Order Formulas on Graph
  Classes of Bounded Expansion
  \thanks{\,StKr and RoRa are supported by the European Research Council
    (ERC) under the European Union's Horizon 2020 research and innovation
    programme (ERC consolidator grant DISTRUCT, agreement No.\ 648527).
    MiPi and SeSi are supported by the National Science Centre of Poland
    via POLONEZ grant agreement UMO-2015/19/P/ST6/03998. This project has
    received funding from the European Union's Horizon 2020 research and
    innovation programme under the Marie Sk\l odowska-Curie grant agreement
    No.\ 665778. MiPi is supported by Foundation for Polish Science (FNP)
    via the START stipend programme.\newline
    \mbox{\qquad}Part of the research for this paper was carried out during
    a stay of JvdH and DAQu at the Logic and Semantics Research Group of TU
    Berlin. JvdH and DAQu would like to thank the group for its
    hospitality. }}

\begin{document}

\maketitle

\begin{abstract}
  \noindent
  A successor-invariant first-order formula is a formula that has access to
  an auxiliary successor relation on a structure's universe, but the model
  relation is independent of the particular interpretation of this
  relation. It is well known that successor-invariant formulas are more
  expressive on finite structures than plain first-order formulas without a
  successor relation. This naturally raises the question whether this
  increase in expressive power comes at an extra cost to solve the
  model-checking problem, that is, the problem to decide whether a given
  structure together with some (and hence every) successor relation is a
  model of a given formula.

  It was shown earlier that adding successor-invariance to first-order
  logic essentially comes at no extra cost for the model-checking problem
  on classes of finite structures whose underlying Gaifman graph is
  planar~\cite{engelmann2012first}, excludes a fixed
  minor~\cite{eickmeyer2013model} or a fixed topological
  minor~\cite{eickmeyer2016model,KreutzerPRS16}. In this work we show that
  the model-checking problem for successor-invariant formulas is
  fixed-parameter tractable on any class of finite structures whose
  underlying Gaifman graphs form a class of bounded expansion. Our result
  generalises all earlier results and comes close to the best tractability
  results on nowhere dense classes of graphs currently known for plain
  first-order logic.

\end{abstract}

\vspace{-0.5cm}
\begin{picture}(0,0) \put(395,-45)
  {\hbox{\includegraphics[scale=0.25]{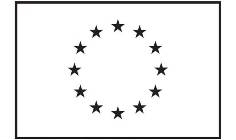}}}
\end{picture} 

\clearpage

\section{Introduction}
\label{sec:intro}

Pinpointing the exact complexity of first-order model-checking has been the
object of a large body of research. The model-checking problem for
first-order logic, denoted $\MC(\FO)$, is the problem of deciding for a
given finite structure~$\strA$ and a formula $\phi\in\FO$ whether
$\strA\models\phi$. Initially, Vardi proposed to distinguish the complexity
of $\MC(\FO)$ into \emph{data}, \emph{formula}, and \emph{combined}
complexity. As shown by Vardi \cite{Vardi82}, for any fixed formula $\phi$
the model checking problem is solvable in polynomial time, i.e.\ the data
complexity of $\MC(\FO)$ is in \textsc{Ptime}. On the other hand, it is
also well known that the general model-checking problem for first-order
logic, that is, with both the formula and the structure regarded as parts
of the input, is \textsc{Pspace}-complete already on a fixed
\mbox{2-element} structure \cite{ChandraM77}.

A more fine-grained analysis of model-checking complexity can be achieved
through the lens of parametrised complexity. In this framework, the
model-checking problem $\MC(\LLL)$ for a logic $\LLL$ is said to be
\emph{fixed-parameter tractable} if it can be solved in time
$f(|\phi|)\cdot|\strA|^c$, for some function $f$ (usually required to be
computable) and a constant $c$ independent of~$\phi$ and $\strA$. The
complexity class FPT of all fixed-parameter tractable problems is the
parametrised analogue to \textsc{Ptime} as model of efficient solvability.
Hence, parametrised complexity lies somewhere between data and combined
complexity, in that the formula is not taken to be fixed and yet has a
different influence on the complexity than the structure. In particular, in
the framework of parametrised complexity the complexity of first-order
model checking on a specific class $\CCC$ of structures can be studied in a
meaningful way. We will denote $\MC(\FO)$ restricted to a class $\CCC$ as
$\MC(\FO,\CCC)$.

In general, a parametrised algorithmic problem takes as input a pair
$(x,k)$, where $x$ is an instance and $k$ is an integer parameter. The
problem is said to be \emph{fixed-parameter tractable} (\emph{FPT} for
short) if it can be solved in time $f(k)\cdot|x|^c$, where~$f$ and $c$ are
as before.

Perhaps the most famous result on the parametrised complexity of
model-checking is Courcelle's theorem \cite{Courcelle90}, which states that
every algorithmic property on graphs definable in \emph{monadic
  second-order logic} can be evaluated in linear time on any class of
graphs of bounded treewidth. An equivalent statement is that
$\MC(\mathrm{MSO},\CCC)$ is FPT via a linear-time algorithm for any class
$\CCC$ of bounded treewidth. Starting with this foundational result, much
work has gone into understanding the complexity of first-order and monadic
second-order model-checking with respect to specific classes of graphs or
structures. In particular, much of this effort has concentrated on sparse
classes of graphs such as planar graphs, graphs of bounded treewidth,
graphs of bounded maximum degree, or classes that exclude a fixed
(topological) minor. Recently, more abstract notions of sparsity have been
considered, namely classes of bounded expansion and nowhere dense classes.
Sparse classes of graphs in this sense have in common that if a class
$\CCC$ is sparse and we close it under taking subgraphs, then it is still
sparse, i.e.\ sparse graphs have no dense subgraphs.

It was shown in \cite{KreutzerT10,KreutzerT10b} that Courcelle's theorem
cannot be extended in full generality much beyond bounded treewidth. For
first-order logic, however, Seese \cite{Seese96} proved that first-order
model checking is fixed-parameter tractable on any class of graphs of
bounded maximum degree. This result was the starting point of a long series
of papers establishing tractability results for first-order model-checking
on sparse classes of graphs, see e.g.
\cite{DawarGK07,dvovrak2013testing,FlumFG02,FrickG01,grohe2014deciding},
and see \cite{grohe2011methods} for a survey. This line of research
culminated in the theorem of Grohe et al.\ \cite{grohe2014deciding} stating
that for any class $\CCC$ of graphs that is closed under taking subgraphs,
$\MC(\FO,\CCC)\in\FPT$ if and only if $\CCC$ is nowhere dense. For sparse
classes of graphs that are closed under taking subgraphs, this yields a
precise characterisation of tractability for first-order model-checking.

The immediate follow-up question is whether this result can be extended in
various ways. One line of research tries to extend tractability to even
more general or different classes of graphs and structures, see e.g.
\cite{GajarskyHLOORS15,GajarskyHOLR16,GanianKOST15}.

In this paper we follow a different route and investigate whether
tractability on sparse classes of graphs can be achieved for more
expressive logics than \FO. It is well-known and easy to see that following
the common approach in finite model theory to add fixed-point or
reachability operators to~\FO\ very quickly results in logics that are not
fixed-parameter tractable (under a standard complexity theoretic assumption
from parametrised complexity theory), even on planar graphs. Therefore, in
this paper we study another classical type of extensions of first-order
logic, namely \emph{successor-} and \emph{order-invariant} first-order
logic.

Over graphs, an \emph{order-invariant} first-order formula is a formula
that, in addition to the edge relation, has access to a linear ordering on
the vertex set of the input graph. However, the formula is required to be
\emph{invariant} under the precise linear ordering chosen. That is, if the
formula is true in a graph~$G$ with a linear order $<$, then it must be
true in $G$ for all choices of linear orderings on $V(G)$. See
\cref{sec:mc} for details. A formula is \emph{successor-invariant} if the
same condition is true for a successor relation instead of a linear
ordering.

Successor- and order-invariant first-order logic have both been studied
intensively in the literature, see e.g.
\cite{BenediktSeg2009,elberfeld2016order,Otto00,potthoff94,Rossman03,rossman2007successor}.
However, the difference between the expressive powers of order-invariant,
successor-invariant, and plain \FO\ on various classes of structures
remains largely unexplored. An unpublished result of Gurevich states that
the expressive power of order-invariant~\FO\ is stronger than that of plain
\FO. Rossman \cite{rossman2007successor} proved that
successor-invariant~\FO\ is more expressive than plain first-order logic.
The construction of \cite{rossman2007successor} creates dense instances
though, and no separation between successor-invariant \FO\ and plain \FO\
is known on sparse classes, say of bounded expansion. On the other hand,
collapse results in this context are known only for very restricted
settings. It is known that order-invariant~\FO\ collapses to plain \FO\ on
trees \cite{BenediktSeg2009,Niemistro05} and on graphs of bounded
treedepth~\cite{EickmeyerEH14}. Moreover, order-invariant \FO\ is a subset
of~MSO on graphs of bounded degree and on graphs of bounded treewidth
\cite{BenediktSeg2009}, and more generally, on decomposable graphs in the
sense of~\cite{elberfeld2016order}.

In \cite{engelmann2012first}, Engelmann et al.\ study the evaluation
complexity of successor and order-invariant first-order logic. They showed
that successor-invariant \FO\ is fixed-parameter tractable on planar
graphs. This was later generalised in \cite{eickmeyer2013model} to classes
of graphs excluding a fixed minor, and then again to classes of graphs
excluding a fixed topological minor in \cite{eickmeyer2016model}. See also
\cite{KreutzerPRS16} for an independent and different proof of this latter
result.

\paragraph*{Our contribution.}
In this paper we narrow the gap between the known tractability results for
plain first-order logic and successor-invariant first-order logic. In
particular, we show that model-checking successor-invariant \FO\ is
fixed-parameter tractable on any class of graphs of bounded expansion.
Classes of bounded expansion generalise classes with excluded topological
minors, and form a natural meta-class one step below nowhere dense classes
of graphs. Thus our result generalises the previous model-checking results
of
\cite{eickmeyer2016model,eickmeyer2013model,engelmann2012first,KreutzerPRS16}.

More precisely, we show that if $\CCC$ is a class of structures of bounded
expansion, then model-checking for successor-invariant first-order formulas
on $\CCC$ can be solved in time $f(|\phi|)\cdot n\cdot\alpha(n)$, where $n$
is the size of the universe of the given structure, $f$ is some function,
and $\alpha(\cdot)$ is the inverse Ackermann function. Note that
model-checking for plain first-order logic can be done in linear time on
classes of bounded expansion \cite{dvovrak2013testing}, thus the running
time of our algorithm is very close to the best known results for plain
\FO. See \cref{thm:main} in \cref{sec:mc} for a precise statement of our
main result.

The natural way of proving tractability for successor-invariant \FO\ on a
specific class $\mathcal{C}$ of graphs is to show that given any graph
$G\in\CCC$, it can be augmented by a new set $F$ of coloured edges such
that a) in $(G, F)$ a successor-relation is first-order definable and b)
$G+F$ falls within a class $\mathcal{D}$ of graphs on which plain
first-order logic is tractable. In this way, model-checking for
successor-invariant \FO\ on the class $\CCC$ is reduced to the
model-checking problem for \FO\ on~$\mathcal{D}$. This technique was
employed in
\cite{eickmeyer2016model,eickmeyer2013model,engelmann2012first,KreutzerPRS16}.
The main problem is how to construct the set of augmentation edges $F$. In
\cite{eickmeyer2016model,eickmeyer2013model,engelmann2012first,KreutzerPRS16}
the authors used topological arguments, based on Robertson and Seymour's
structure theorem for classes with excluded
minors~\cite{robertsonseymour1999} or its generalisation by Grohe and Marx
to classes with excluded topological minors~\cite{grohemarx2015}.

For classes of bounded expansion, the object of study in this paper, no
such topological methods exist. Instead we rely on a characterisation of
bounded expansion classes by \emph{generalised colouring numbers}. The
definition of these graph parameters is roughly based on measuring
reachability properties in a linear vertex ordering of the input graph. Any
such ordering yields a very weak form of decomposition of a graph in terms
of an elimination tree. The main technical contribution of this paper is
that we find a way to control these elimination trees so that we can use
them to define a set $F$ of new edges with the following properties: a) $F$
forms a spanning tree of the input graph $G$, b) $F$ has maximum degree at
most $3$, and c) after adding all the edges of $F$ to the graph, the
colouring numbers are still bounded. See \cref{thm:main-technical} in
\cref{sec:mc} for a formal statement of this main technical contribution.

This construction, besides its use in this paper, yields a new insight into
the elimination trees generated by colouring numbers. We believe it may
prove useful for future research as well.

\paragraph*{Organisation.}
In \cref{sec:prelims} we fix the terminology and notation used throughout
the paper and recall the notions from the theory of sparse graphs, in
particular the generalised colouring numbers. In \cref{sec:mc} we show how
having access to a low degree spanning tree in a graph can be used to
reduce model-checking for successor-invariant \FO\ to model-checking for
plain \FO, thus proving our main result. Finally, in \cref{sec:tree} we
present out main technical contribution: the construction of a low degree
spanning tree that can be added to a graph without increasing the colouring
numbers too much.

\section{Preliminaries}\label{sec:prelims}

\paragraph*{Notation.}
By $\N$ we denote the set of nonnegative integers. For a set $X$, by
$\binom{X}{2}$ we denote the set of unordered pairs of elements of $X$,
that is, $2$-element subsets of $X$. By $\alpha(\cdot)$ we denote the
inverse Ackermann function.

We use standard graph-theoretical notation; see e.g.\
\cite{diestel2012graph} for reference. All graphs considered in this paper
are finite, simple, and undirected. For a graph $G$, by $V(G)$ and
$E(G)\subseteq \binom{V(G)}{2}$ we denote the vertex and edge sets of $G$,
respectively. For a vertex $v$ and an edge $e$, we write $v\in G$ and
$e\in G$ meaning $v\in V(G)$ and $e\in E(G)$, respectively. A graph~$H$ is
a \emph{subgraph} of~$G$ if $V(H)\subseteq V(G)$ and $E(H)\subseteq E(G)$.
For a vertex subset $X\subseteq V(G)$, the \emph{subgraph induced by $X$}
is equal to $G[X]=\bigl(X,E(G)\cap\binom{X}{2}\bigr)$. For a vertex $v$, we
write $G-v$ for $G[V(G)\setminus\{v\}]$. For a set of unordered pairs
$F\subseteq\binom{V(G)}{2}$, by $G+F$ we denote the graph
$(V(G),E(G)\cup F)$.

For a nonnegative integer $\ell$, a \emph{walk of length $\ell$} in~$G$ is
a sequence $P=(v_1,\ldots, v_{\ell+1})$ of vertices such that
$v_iv_{i+1}\in E(G)$ for all $1\le i\le \ell$. Vertices $v_i$ and edges
$v_iv_{i+1}$ are \emph{traversed} by the walk~$P$, $v_1$ and $v_{\ell+1}$
are the \emph{endpoints} of $P$, while $v_2,\ldots,v_{\ell}$ are the
\emph{internal} vertices of~$P$. A walk $P$ is a \emph{path} if the
vertices traversed by it are pairwise different. A walk or path
\emph{connects} its endpoints.

A graph $G$ is \emph{connected} if any pair of its vertices can be
connected by a path. The \emph{distance} between vertices $u,v\in V(G)$,
denoted $\dist_G(u,v)$, is the minimum length of a path between $u$ and $v$
in $G$. The \emph{radius} of a connected graph $G$ is defined as
$\min\limits_{u\in V(G)}\max\limits_{v\in V(G)}\dist_G(u,v)$.

A graph $T$ is a \emph{tree} if it is connected and has no cycles;
equivalently, it is connected and has exactly $|V(T)|-1$ edges.
\emph{Rooting} a tree $T$ in some vertex $w\in V(T)$ imposes
\emph{child-parent} and \emph{ancestor-descendant} relations in $T$. More
precisely, a vertex $v$ is an \emph{ancestor} of a vertex $u$ if it lies on
the unique path from $u$ to the root $w$, and it is the \emph{parent} of
$u$ if it is the immediate successor of $u$ on this path. Thus, each vertex
is both an ancestor and a descendant of itself. We say \emph{strict
  ancestor or descendant} to express that the considered vertices are
different.

\paragraph*{Shallow minors and bounded expansion.}
A graph $H$ is a \emph{minor} of $G$, written $H\minor G$, if there are
pairwise disjoint connected subgraphs $(I_u)_{u\in V(H)}$ of $G$, called
\emph{branch sets}, such that whenever $uv\in E(H)$, then there are
$x_u\in I_u$ and $x_v\in I_v$ with $x_ux_v\in E(G)$. We call the family
$(I_u)_{u\in V(H)}$ a \emph{minor model} of $H$ in~$G$. A graph $H$ is a
\emph{depth-$r$ minor} of~$G$, denoted $H\minor_r G$, if there is a minor
model $(I_u)_{u\in V(H)}$ of~$H$ in $G$ such that each subgraph $I_u$ has
radius at most $r$.

For a graph $H$, we write $d(H)$ for the \emph{average degree} of~$H$, that
is, for the number $2|E(H)|/|V(H)|$. A class of graphs $\CCC$ has
\emph{bounded expansion} if there is a function $f\colon\N\rightarrow\N$
such that for all nonnegative integers $r$, we have $d(H)\le f(r)$ for
every $H\minor_r G$ with $G\in\CCC$.

\paragraph*{Generalised colouring numbers.}
In this paper we will not rely on the above, original, definition of
classes of bounded expansion, but on their alternative characterisation via
\emph{generalised colouring numbers}. Let us fix a graph $G$. By $\Pi(G)$
we denote the set of all linear orderings of~$V(G)$. For $L\in\Pi(G)$, we
write $u<_L v$ if $u$ is smaller than $v$ in $L$, and $u\le_L v$ if
$u<_L v$ or $u=v$.

For a nonnegative integer $r$, we say that a vertex $u$ is \emph{strongly
  $r$-reachable} from a vertex~$v$ with respect to~$L$, if $u\le_L v$ and
there is a path $P$ of length at most $r$ that starts in $v$, ends in~$u$,
and all its internal vertices are larger than $v$ in $L$. By
$\Sreach_r[G,L,v]$ we denote the set of vertices that are strongly
$r$-reachable from~$v$ with respect to $L$. Note that
$v\in\Sreach_r[G,L,v]$ for any vertex~$v$.

We define the \emph{$r$-colouring number} of $G$ (with respect to $L$) as
follows:
\[\col_r(G,L)=\max_{v\in V(G)} \abs{\Sreach_r[G,L,v]}\qquad
\text{and}\qquad \col_r(G)=\min_{L\in \Pi(G)} \col_r(G,L).\]

For a nonnegative integer $r$ and ordering $L\in\Pi(G)$, the
\mbox{\emph{$r$-admissibility}} $\adm_r[G,L,u]$ of a vertex $v$ with
respect to $L$ is defined as the maximum size of a family $\Pp$ of paths
that satisfies the following two properties:

\qitem{$\bullet$}{each path $P\in\Pp$ has length at most $r$, starts in
  $v$, ends in a vertex that is smaller than~$v$ in~$L$, and all its
  internal vertices are larger than~$v$ in $L$;}

\qitem{$\bullet$}{the paths in $\Pp$ are pairwise vertex-disjoint, apart
  from sharing the start vertex $v$.}

\smallskip\noindent
The \emph{$r$-admissibility} of $G$ (with respect to $L$) is defined
similarly to the $r$-colouring number:
\[\adm_r(G,L)=\max_{v\in V(G)} \adm_r[G,L,v]\qquad \text{and}\qquad
\adm_r(G)=\min_{L\in \Pi(G)} \adm_r(G,L).\]

The $r$-colouring numbers were introduced by Kierstead and Yang
\cite{kierstead2003orders}, while $r$-admissibility was first studied by
Dvo\v{r}\'ak~\cite{dvovrak13}. It was shown that those parameters are
related as follows.

\begin{lemma}[\rm Dvo\v{r}\'ak
  \cite{dvovrak13}]\label{lem:gen-col-ineq}\mbox{}\\*
  For each graph $G$, nonnegative integer $r$, and vertex ordering
  $L\in\Pi(G)$, we have
  \[\adm_r(G,L)\le \col_r(G,L)\le (\adm_r(G,L))^r.\]
\end{lemma}

\noindent
We remark that in Dvo\v{r}\'ak's work, the reachability sets never 
include the starting vertex, hence the above inequality is stated
slightly different in \cite{dvovrak13}.

As proved by Zhu \cite{zhu2009coloring}, the generalised colouring numbers
are tightly related to densities of low-depth minors, and hence they can be
used to characterise classes of bounded expansion.

\begin{theorem}[\rm Zhu
  \cite{zhu2009coloring}]\label{thm:b_exp_b_deg}\mbox{}\\*
  A class $\CCC$ of graphs has bounded expansion if and only if there is a
  function $f\colon \N\rightarrow\N$ such that $\col_r(G)\le f(r)$ for all
  $r\in\N$ and all $G\in\CCC$.
\end{theorem}

\noindent
By \cref{lem:gen-col-ineq}, we may equivalently demand that there is a
function $f\colon \N\rightarrow\N$ such that $\adm_r(G)\le f(r)$ for all
nonnegative integers $r$ and all $G\in\CCC$.

As shown by Dvo\v{r}\'ak \cite{dvovrak13}, on classes of bounded expansion
one can compute $\adm_r(G)$ in linear fixed-parameter time, parametrised by
$r$. More precisely, we have the following.

\begin{theorem}[\rm Dvo\v{r}\'ak
  \cite{dvovrak13}]\label{thm:adm-compute}\mbox{}\\*
  Let $\CCC$ be a class of bounded expansion. Then there is an algorithm
  that, given a graph $G\in \CCC$ and a nonnegative integer $r$, computes a
  vertex ordering $L\in \Pi(G)$ with $\adm_r(G,L)=\adm_r(G)$ in time
  $f(r)\cdot |V(G)|$, for some computable function $f$.
\end{theorem}

\noindent
We remark that Dvo\v{r}\'ak states the result in \cite{dvovrak13} as the
existence of a linear-time algorithm for each fixed value of~$r$. However,
an inspection of the proof reveals that it is actually a single
fixed-parameter algorithm that can take $r$ as input. To the best of our
knowledge, a similar result for computing $\col_r(G)$ is not known, but by
\cref{lem:gen-col-ineq} we can use admissibility to obtain an approximation
of the $r$-colouring number of a given graph from a class of bounded
expansion.

\section{Model-checking}\label{sec:mc}

We start by introducing successor-invariant first-order formulas and
stating the main theorem (\cref{thm:main}) formally. Next, we show how to
prove this theorem assuming our main technical result,
\cref{thm:main-technical}. The proof follows by a combination of several
tools borrowed from the literature.

\paragraph*{Successor-invariant first-order formulas.}
A \emph{finite and purely relational signature} $\tau$ is a finite set
$\{R_1,\ldots,R_k\}$ of relation symbols, where each relation symbol~$R_i$
has an associated arity $a_i$. A \emph{finite $\tau$-structure} $\strA$
consists of a finite set~$A$ (the \emph{universe} of $\strA$) and a
relation $R_i(\strA)\subseteq A^{a_i}$ for each relation symbol
$R_i\in \tau$. If $ \strA$ is a finite $\tau$-structure, then the
\emph{Gaifman graph} of $\strA$, denoted $G(\strA)$, is the graph on the
vertex set $A$ in which two elements $u,v\in A$ are adjacent if and only if
$u\neq v$ and $u$ and $v$ appear together in some relation $R_i(\strA)$ of
$\strA$. We say that a class $\CCC$ of finite $\tau$-structures has
\emph{bounded expansion} if the graph class 
$G(\CCC)=\{G(\strA)\mid \strA\in\CCC\}$ has bounded expansion. Similarly,
for $r\in\N$, we write $\adm_r(\strA)$ for $\adm_r(G(\strA))$, etc.

Let $V$ be a set. A \emph{successor relation} on $V$ is a binary relation
$S\subseteq V\times V$ such that $(V,S)$ is a directed path of length
$|V|-1$. Let $\tau$ be a finite relational signature. A formula
$\phi\in\FO[\tau\cup\{S\}]$ is \emph{successor-invariant} if for all
$\tau$-structures $\strA$ and for all successor relations $S_1,S_2$ on
$V(\strA)$ it holds that
$(\strA,S_1)\models\phi\Longleftrightarrow(\strA,S_2)\models\phi$. We
denote the set of all such successor-invariant first-order formulas by
$\FOs$. Note that the set $\FOs$ is not decidable, and hence one usually
does not speak of successor-invariant first-order logic, as for a logic one
usually requires a decidable syntax \cite{gurevich1985logic}. For any
$\phi\in\FOs$ and any $\tau$-structure $\strA$, we denote
$\strA\models_{\mathrm{succ-inv}}\phi$ if for any (equivalently, every)
successor relation $S$ on the universe of $\strA$ it holds that
$(\strA,S)\models \phi$.

With these definitions in mind, we can finally state our main result
formally.

\begin{theorem}\label{thm:main}\mbox{}\\*
  Let $\tau$ be a finite and purely relational signature and let $\CCC$ be
  a class of $\tau$-structures of bounded expansion. Then there exists an
  algorithm that, given a finite $\tau$-structure $\strA\in\CCC$ and a
  formula $\phi\in \FOs$, verifies whether
  $\strA\models_{\mathrm{succ-inv}}\phi$ in time $f(|\phi|)\cdot n\cdot
  \alpha(n)$, where $f$ is a function and~$n$ is the size of the universe
  of $\strA$.
\end{theorem}

\noindent
In the language of parametrised complexity, \cref{thm:main} essentially
states that the model-checking problem for successor-invariant first-order
formulas is fixed-parameter tractable on classes of finite structures whose
underlying Gaifman graph belongs to a fixed class of bounded expansion.
There is a minor caveat, though. The formal definition of fixed-parameter
tractability, see e.g.~\cite{flum2006parameterized}, requires the function
$f$ to be computable, which is not asserted by \cref{thm:main}. In order to
have this property, it suffices to assume that the class $\CCC$ is
\emph{effectively of bounded expansion}. In the characterisation of
\cref{thm:b_exp_b_deg}, this means that there exist a \emph{computable}
function $f:\N\rightarrow\N$ such that $\col_r(\strA)\leq f(r)$ for each
$\strA\in \CCC$. See \cite{grohe2014deciding} for a similar discussion
regarding model-checking first-order logic on (effectively) nowhere dense
classes of graphs.

As we mentioned in \cref{sec:intro}, fixed-parameter tractability of
model-checking successor-invariant \FO\ has been shown earlier for planar
graphs \cite{engelmann2012first}, graphs excluding a fixed
minor~\cite{eickmeyer2013model}, and graphs excluding a fixed topological
minor \cite{eickmeyer2016model,KreutzerPRS16}. As all the above-mentioned
classes have (effectively) bounded expansion, \cref{thm:main} thus
generalises all the previously known results in this area. Let us remark
that on general relational structures, the model-checking problem for plain
first-order logic is complete for the parametrised complexity class
$\mathrm{AW}[\star]$, and hence unlikely to be fixed-parameter tractable
\cite{flum2006parameterized}.

\paragraph*{From a spanning tree to a successor relation.}
In principle, our approach follows that of all earlier results on
successor-invariant model-checking. As we would like to check whether
$\strA\models_{\mathrm{succ-inv}}\phi$, we may compute an arbitrary
successor relation $S$ on the universe of $\strA$, and verify whether
$(\strA,S)\models \phi$. Of course, we will try to compute a successor
relation $S$ so that adding it to $\strA$ preserves the structural
properties as much as possible, so that model-checking on~$(\strA,S)$ can
be done efficiently. Ideally, if $G(\strA)$ contained a Hamiltonian path,
we could add a successor relation without introducing any new edges to
$G(\strA)$. However, in general this might be impossible.

The other helpful ingredient is that we do not actually have to add a
successor relation, but it suffices to add some structural information so
that a first-order formula can interpret a successor relation. This
approach is known as the interpretation method \cite{grohe2011methods} and
can be used to reduce successor-invariant model-checking to the plain
first-order case. In our concrete case, Eickmeyer et al.\
\cite{eickmeyer2013model} have shown that adding a spanning tree of
constant maximum degree is enough to be able to interpret some successor
relation. Here, for relational signatures $\tau'\supseteq \tau$, a
$\tau'$-structure~$\strA'$ is a \emph{$\tau'$-expansion} of a
$\tau$-structure $\strA$ if after dropping relations $R(\strA')$ for
relation symbols $R\in \tau'\setminus \tau$, $\strA'$ becomes equal to
$\strA$.

\begin{lemma}[\rm Lemma 4.4 of Eickmeyer et al.\ \cite{eickmeyer2013model},
  adjusted]\label{thm:eickmeyer-combine}\mbox{}\\*
  Let $\tau$ be a finite and purely relational signature, and let $k$ be a
  positive integer. Suppose we are given a finite $\tau$-structure~$\strA$,
  together with a spanning tree $T$ of $G(\strA)$ with maximum degree at
  most~$k$. Then there is a finite and purely relational signature $\tau_k$
  and a first-order formula $\psi^{(k)}_{\mathrm{succ}}(x, y)$, both
  depending only on $k$, and a $(\tau \cup \tau_k)$-expansion $\strA'$ of
  $\strA$, such that

  \smallskip\qitem{$\bullet$}{the Gaifman graphs of $\strA'$ and $\strA$
    are equal;}

  \smallskip\qitem{$\bullet$}{$\psi^{(k)}_{\mathrm{succ}}(x, y)$ defines a
    successor relation on $\strA'$.}

  \smallskip\noindent
  Moreover, for a fixed $k$ and given $\strA$ and $T$, one can
  compute~$\strA'$ in time linear in the size of the input.
\end{lemma}

\noindent
Informally speaking, the structure~$\tau'$ contains the edges that form the
spanning tree, using a new colour. Important for us is that the
construction guarantees that the Gaifman graphs of $\strA$ and $\strA'$ are
the same, and thus so are their structural properties.

\cref{thm:eickmeyer-combine} differs from the original statement of
Eickmeyer et al.\ \cite{eickmeyer2013model} in two ways. First, Eickmeyer
et al.\ \cite{eickmeyer2013model} state the running time only as
polynomial, however a verification of the proof yields that the
straightforward implementation runs in linear time. Second, Eickmeyer et
al.\ \cite{eickmeyer2013model} require the existence of a $k$-walk instead
of a spanning tree of maximum degree~$k$. Here, a \emph{$k$-walk} is a walk
in the graph that visits every vertex at least once and at most $k$ times.
Observe that if a graph has a spanning tree of maximum degree at most $k$,
then in particular it contains a $k$-walk that can be computed from the
given spanning tree in polynomial time by performing a depth-first search
on the tree. Thus, the assumption of having a spanning tree of maximum
degree $k$ is sufficient for \cref{thm:eickmeyer-combine} to work. We find
working with spanning trees more natural than with $k$-walks, however both
approaches are essentially equivalent: if a graph admits a $k$-walk, then
it has a spanning tree of maximum degree at most $2k$.

As discussed above, \cref{thm:eickmeyer-combine} essentially reduces
model-checking successor-invariant first-order formulas to plain
first-order logic, provided we can expose some spanning tree with maximum
degree bounded by a constant in the Gaifman graph of the given structure.
In general this might be not possible; e.g.\ the Gaifman graph could be a
star. Therefore, the idea is to add a carefully constructed binary relation
to the structure so that such a low-degree spanning tree can be found,
while maintaining the property that the structure still belongs to a class
where model-checking first-order logic is fixed-parameter tractable. This
approach was used in all the previous works
\cite{eickmeyer2016model,eickmeyer2013model,KreutzerPRS16}, and, with a
small twist, we will use it also here. More precisely, in the next section
we will prove the following theorem which gives a construction of a
low-degree spanning tree that can be added to a graph without increasing
its colouring numbers too much.

\begin{theorem}\label{thm:main-technical}\mbox{}\\*
  There exists an algorithm that, given a graph $G$ and an ordering $L$ of
  $V(G)$, computes a set of unordered pairs $F\subseteq\binom{V(G)}{2}$
  such that the graph $T=(V(G),F)$ is a tree of maximum degree at most $3$
  and
  \[\adm_r(G+F,L)\le 2+3\cdot\col_{2r}(G,L).\]
  The running time of the algorithm is $\Oh((m+n)\cdot \alpha(m))$, where
  $m=|E(G)|$ and $n=|V(G)|$.
\end{theorem}

\noindent
We remark that in the earlier work
\cite{eickmeyer2016model,eickmeyer2013model,KreutzerPRS16}, the bound on
the maximum degree of the constructed spanning tree was a constant
depending on the size of the excluded (topological) minor, while
\cref{thm:main-technical} always bounds the maximum degree of the spanning
tree by $3$.

\paragraph*{Model-checking plain FO.}
Before showing how \cref{thm:main} follows from \cref{thm:main-technical},
we first need to draw upon the literature on model-checking first-order
logic on classes of bounded expansion. The following statement encapsulates
the model-checking results of Dvo\v{r}\'ak et al.\
\cite{dvovrak2013testing} and of Grohe and Kreutzer
\cite{grohe2011methods}. However, it is slightly stronger than the
statements claimed in \cite{dvovrak2013testing,grohe2011methods}. We will
later argue how this statement follows from the approach presented in these
works.

\begin{theorem}\label{thm:plain-FO-MC}\mbox{}\\*
  Let $\tau$ be a finite and purely relational signature. Then for every
  formula $\phi\in \FO[\tau]$ there exists a nonnegative integer $r(\phi)$,
  computable from $\phi$, such that the following holds. Given a
  $\tau$-structure~$\strA$, it can be verified whether $\strA\models \phi$
  in time $f(|\phi|,\col_{r(\phi)}(\strA))\cdot n$, where $n$ is the size
  of the universe of $\strA$ and $f$ is a computable function.
\end{theorem}

\noindent
Observe that if $\strA$ is drawn from a fixed class of bounded expansion
$\CCC$, then $\adm_{r(\phi)}(\strA)$ is a parameter depending only on
$\phi$, hence we recover fixed-parameter tractability of
model-checking~\FO\ on any class of bounded expansion, parametrised by the
length of the formula. \cref{thm:plain-FO-MC} is stronger than this latter
statement in that it says that the input structure does not need to be
drawn from a fixed class of bounded expansion, where the colouring number
is bounded in terms of the radius~$r$ for all values of $r$, but it
suffices to have a bound on the colouring numbers up to some radius
$r(\phi)$, which depends only on the formula $\phi$. We need this
strengthening in our algorithm for the following reason: When adding a
low-degree spanning tree to the Gaifman graph, we are not able to control
all the colouring numbers at once, but only for some particular value of
the radius. \cref{thm:plain-FO-MC} ensures that this is sufficient for the
model-checking problem to remain tractable.

We now sketch how \cref{thm:plain-FO-MC} may be derived from the works of
Dvo\v{r}\'ak et al.\ \cite{dvovrak2013testing} and of Grohe and
Kreutzer~\cite{grohe2011methods}. We prefer to work with the algorithm of
Grohe and Kreutzer~\cite{grohe2011methods}, because we find it conceptually
simpler. For a given quantifier rank $q$ and an nonnegative integer $i\le
q$, the algorithm computes the set of all \emph{types} $\mathfrak{R}_i^q$
realised by $i$-tuples in the input structure $\strA$: for a given
$i$-tuple of elements $\overline{a}$, its \emph{type} is the set of all
\FO\ formulas $\phi(\overline{x})$ with~$i$ free variables and quantifier
rank at most $q-i$ for which $\phi(\overline{a})$ holds. Note that for
$i=0$, this corresponds to the set of sentences of quantifier rank at most
$q$ that hold in the structure, from which the answer to the model-checking
problem can be directly read; whereas for $i=q$, we consider
quantifier-free formulas with $q$ free variables. Essentially,
$\mathfrak{R}_q^q$ is computed explicitly, and then one inductively
computes $\mathfrak{R}_i^q$ based on $\mathfrak{R}_{i+1}^q$. The above
description is, however, a bit too simplified, as each step of the
inductive computation introduces new relations to the structure, but does
not change its Gaifman graph; we will explain this later.

When implementing the above strategy, the assumption that the structure is
drawn from a class of bounded expansion is used via \emph{treedepth-$p$
  colourings}, a colouring notion functionally equivalent to the
generalised colouring numbers. More precisely, a treedepth-$p$ colouring of
a graph $G$ is a colouring $\gamma:V(G)\rightarrow\Gamma$, where $\Gamma$
is a set of colours, such that for any subset $C\subseteq\Gamma$ of~$i$
colours, $i\le p$, the vertices with colours from~$C$ induce a subgraph of
treedepth at most~$i$. The \emph{treedepth-$p$ chromatic number} of a graph
$G$, denoted $\chi_p(G)$, is the smallest number of colours $|\Gamma|$
needed for a \mbox{treedepth-$p$} colouring of~$G$. As proved by Zhu
\cite{zhu2009coloring}, the treedepth-$p$ chromatic numbers are bounded in
terms of \mbox{$r$-colouring} numbers as follows:

\begin{theorem}[\rm Zhu
  \cite{zhu2009coloring}]\label{lem:low-td-col}\mbox{}\\*
  For any graph $G$ and $p\in \N$ we have
  \[\chi_p(G)\le \bigl(\col_{2^{p-2}}(G)\bigr)^{2^{p-2}}.\]
\end{theorem}

\noindent
Moreover, an appropriate treedepth-$p$ colouring can be constructed in
polynomial time from an ordering $L\in\Pi(G)$, certifying an upper bound on
$\col_{2^{p-2}}(G)$.

The computation of both $\mathfrak{R}_q^q$ and $\mathfrak{R}_i^q$ from
$\mathfrak{R}_{i+1}^q$ is done by rewriting every possible type as a purely
existential formula. Each rewriting step, however, enriches the signature
by unary relations corresponding to colours of some \mbox{treedepth-$p$}
colouring $\gamma$, as well as binary relations representing edges of
appropriate treedepth decompositions certifying that $\gamma$ is correct.
However, the binary relations are added in a way that the Gaifman graph of
the structure remains intact. For us it is important that in all the steps,
the parameter $p$ used for the definition of $\gamma$ depends only on $q$
and $i$ in a computable manner. Thus, by \cref{lem:low-td-col}, to ensure
that $\gamma$ uses a bounded number of colours, we only need to ensure the
boundedness of $\col_{r(q)}(\strA)$ for some computable function $r(q)$. By
taking~$q$ to be the quantifier rank of the input formula, the statement of
\cref{thm:plain-FO-MC} follows.

\paragraph*{From a spanning tree to model-checking.}
We can now combine all the ingredients and show how our main result follows
from \cref{thm:main-technical}.

\begin{proof}[of \cref{thm:main}, assuming \cref{thm:main-technical}]\quad
  Consider the signature $\tau_3$ and the first-order formula
  $\psi^{(3)}_{\mathrm{succ}}(x, y)$ given by \cref{thm:eickmeyer-combine}
  for $k=3$. Let $\tau'=\tau\cup \tau_3\cup \{T\}$, where $T$ is a binary
  relation symbol not used in $\tau\cup \tau_3$. Let $\phi'\in \FO[\tau']$
  be constructed from the input formula $\phi$ by replacing each usage of
  the successor relation $S(x,y)$ with the formula
  $\psi^{(3)}_{\mathrm{succ}}(x, y)$.

  Using \cref{thm:plain-FO-MC} for the signature $\tau'$, compute the value
  of $r=r(\phi')$. Next, using the algorithm of \cref{thm:adm-compute},
  compute a vertex ordering $L\in \Pi(\strA)$ such that
  $\adm_{2r}(\strA,L)=\adm_{2r}(\strA)$. Apply \cref{thm:main-technical} to
  $G(\strA)$ and $L$, thus obtaining a tree $T$ with the universe of
  $\strA$ as the vertex set, such that the maximum degree in $T$ is at most
  $3$ and
  \[\adm_r(\strA_T)\le \adm_{r}(\strA_T,L)\le 2+2\cdot\col_{2r}(\strA,L)\le
  2+2(\adm_{2r}(\strA,L))^{2r}= 2+2(\adm_{2r}(\strA))^{2r}.\]
  Here $\strA_T$ is the $\tau\cup \{T\}$-extension of $\strA$ obtained by
  adding the edges of $T$ as a binary relation. Next, apply the algorithm
  of \cref{thm:eickmeyer-combine} to $\strA_T$ and $k=4$, thus computing a
  $\tau'$-structure~$\strA'$ with the same Gaifman graph as $\strA_T$, in
  which $\psi^{(3)}_{\mathrm{succ}}(x, y)$ defines a successor relation. It
  is then clear that
  \[\strA\models_{\mathrm{succ-inv}}\phi\quad \Longleftrightarrow\quad
  \strA'\models\phi'.\] 
  It remains to apply the algorithm of \cref{thm:plain-FO-MC} to $\strA'$
  and~$\phi'$, which runs in time $f(|\phi|)\cdot n$ due to the bound on
  $\adm_r(\strA_T)=\adm_r(\strA')$. Observe that all the other steps also
  work in time $f(|\phi|)\cdot n$, apart from the application of the
  algorithm of \cref{thm:main-technical}, which takes time
  $\Oh((m+n)\cdot\alpha(m))$, where $m=|E(G(\strA))|$. However, in classes
  of bounded expansion the number of edges is bounded linearly in the
  number of vertices, hence the time complexity analysis follows.
\end{proof}

\section{Constructing a low-degree spanning tree}\label{sec:tree}

\newcommand{\Zup}{Z^{\uparrow}}
\newcommand{\Zdown}{Z^{\downarrow}}
\newcommand{\Fnew}{F_{\textrm{new}}}

In this section we prove \cref{thm:main-technical}. The main step towards
this goal is the corresponding statement for connected graphs, as expressed
in the following lemma.

\begin{lemma}\label{lem:connected}\mbox{}\\*
  There exists an algorithm that, given a connected graph $G$ and a vertex
  ordering $L$ of $G$, computes a set of unordered pairs
  $F\subseteq\binom{V(G)}{2}$ such that the graph $T=(V(G),F)$ is a tree of
  maximum degree at most $3$ and
  \[\adm_r(G+F,L)\le 3\cdot\col_{2r}(G,L).\]
  The running time of the algorithm is $\Oh(m\cdot\alpha(m))$, where
  $m=|E(G)|$.
\end{lemma}

\noindent
We first show that \cref{thm:main-technical} follows easily
from \cref{lem:connected}.

\begin{proof}[of \cref{thm:main-technical}, assuming
  \cref{lem:connected}]\quad Let $G$ be a (possibly disconnected) graph,
  and let $G_1,\ldots,G_p$ be the connected components of $G$. For each
  $i\in\{1,\ldots,p\}$, let $L_i$ be the ordering obtained by restricting
  $L$ to $V(G_i)$. Obviously $\col_{2r}(G_i,L_i)\le \col_{2r}(G,L)$.

  Apply the algorithm of \cref{lem:connected} to $G_i$ and $L_i$, obtaining
  a subset of unordered pairs $F_i$ such that $T_i=(V(G_i),F_i)$ is a tree
  of maximum degree at most $3$ and
  \[\adm_r(G_i+F_i,L_i)\le 3\cdot\col_{2r}(G_i,L_i)\le
  3\cdot \col_{2r}(G,L).\]
  For each $i\in\{1,\ldots,p\}$, select a vertex $v_i$ of $G_i$ with degree
  at most $1$ in $T_i$; since $T_i$ is a tree, such a vertex exists. Define
  \[F= \{v_1v_2,v_2v_3,\ldots,v_{p-1}v_p\}\cup\bigcup_{i=1}^p F_i.\]
  Obviously we have that $T=(V(G),F)$ is a tree. Observe that it has
  maximum degree at most $3$. This is because each vertex $v_i$ had degree
  at most $1$ in its corresponding tree $T_i$, and hence its degree can
  grow to at most $3$ after adding edges $v_{i-1}v_i$ and $v_iv_{i+1}$. By
  \cref{lem:connected}, the construction of $T$ takes time
  $\Oh(m+n\alpha(n))$.

  It remains to show that $\adm_r(G+F,L)\le 2+3\cdot\col_{2r}(G,L)$. Take
  any vertex $u$ of $G$, say $u\in V(G_i)$, and let $\Pp$ be a set of paths
  of length at most $r$ that start in~$u$, are pairwise vertex-disjoint
  (apart from $u$), and end in vertices smaller than~$u$ in~$L$ while
  internally traversing only vertices larger than $u$ in~$L$. Observe that
  at most two of the paths from $\Pp$ can use any of the edges from the set
  $\{v_1v_2,v_2v_3,\ldots,v_{p-1}v_p\}$, since any such path has to use
  either $v_{i-1}v_i$ or $v_iv_{i+1}$. The remaining paths are entirely
  contained in $G_i+F_i$, and hence their number is bounded by
  $\adm_r(G_i+F_i,L_i)\le 3\col_{2r}(G,L)$. The theorem follows.
\end{proof}

\noindent
In the remainder of this section we focus on \cref{lem:connected}.

\begin{proof}[of \cref{lem:connected}]\quad We begin our proof by showing
  how to compute the set $F$. This will be a two step process, starting
  with an \emph{elimination tree}. For a connected graph~$G$ and an
  ordering~$L$ of $V(G)$, we define the \emph{(rooted) elimination tree}
  $S(G,L)$ of $G$ imposed by~$L$ (cf.\
  \cite{bodlaender1998rankings,schaffer1989optimal}) as follows. If
  $V(G)=\{v\}$, then the rooted elimination tree $S(G,L)$ is just the tree
  on the single vertex $v$. Otherwise, the root of $S(G,L)$ is the vertex
  $w$ that is the smallest with respect to the ordering~$L$ in $G$. For
  each connected component $C$ of $G-w$ we construct a rooted elimination
  tree $S(C,L|_{V(C)})$, where $L|_{V(C)}$ denotes the restriction of $L$
  to the vertex set of $C$. These rooted elimination trees are attached
  below $w$ as subtrees by making their roots into children of $w$. Thus,
  the vertex set of the elimination tree $S(G,L)$ is always equal to the
  vertex set of $G$. See \Cref{fig:G_S_U} for an illustration. The solid
  black lines are the edges of $G$; the dashed blue lines are the edges of
  $S$. The ordering $L$ is given by the numbers written in the vertices.

  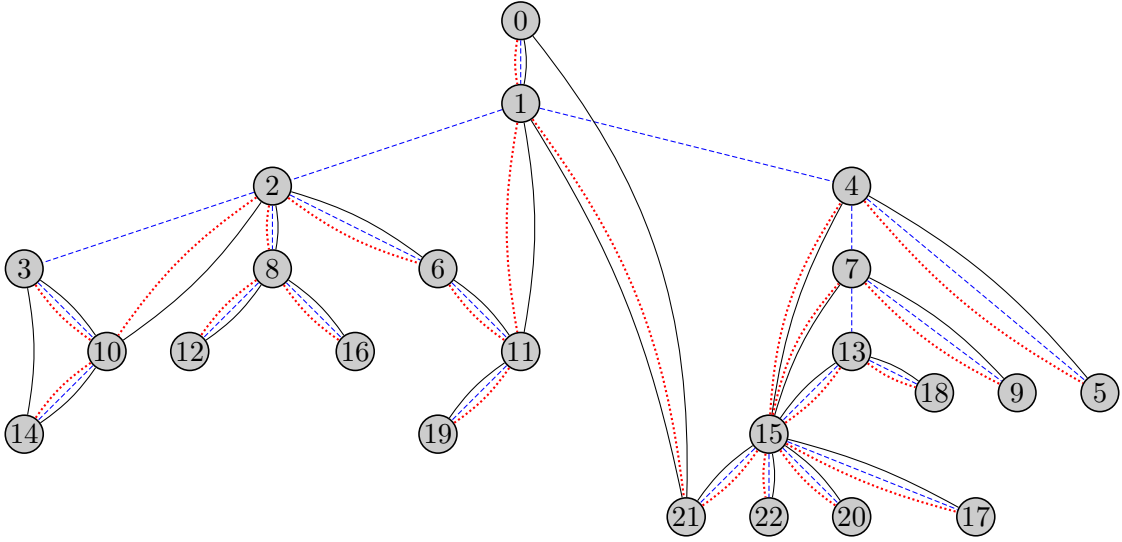
\begin{figure*}
    \centering
    \begin{tikzpicture}[scale=1.1]
      \node[vertex] (0) at (0,0){$0$};
      \node[vertex] (1) at (0,-1){$1$};
      \node[vertex] (2) at (-3,-2){$2$};
      \node[vertex] (3) at (-6,-3){$3$};
      \node[vertex] (4) at (4,-2){$4$};
      \node[vertex] (5) at (7,-4.5){$5$};
      \node[vertex] (6) at (-1,-3){$6$};
      \node[vertex] (7) at (4,-3){$7$};
      \node[vertex] (8) at (-3,-3){$8$};
      \node[vertex] (9) at (6,-4.5){$9$};
      \node[vertex] (14) at (-6,-5){$14$};
      \node[vertex] (19) at (-1,-5){$19$};
      \node[vertex] (12) at (-4,-4){$12$};
      \node[vertex] (13) at (4,-4){$13$};
      \node[vertex] (10) at (-5,-4){$10$};
      \node[vertex] (15) at (3,-5){$15$};
      \node[vertex] (16) at (-2,-4){$16$};
      \node[vertex] (17) at (5.5,-6){$17$};
      \node[vertex] (18) at (5,-4.5){$18$};
      \node[vertex] (11) at (0,-4){$11$};
      \node[vertex] (20) at (4,-6){$20$};
      \node[vertex] (21) at (2,-6){$21$};
      \node[vertex] (22) at (3,-6){$22$};

      \draw[S] (0) -- (1);
      \draw[S] (1) -- (2);
      \draw[S] (1) -- (4);
      \draw[S] (2) -- (3);
      \draw[S] (2) -- (6);
      \draw[S] (4) -- (5);
      \draw[S] (10) -- (14);
      \draw[S] (3) -- (10);
      \draw[S] (2) -- (8);
      \draw[S] (8) -- (12);
      \draw[S] (8) -- (16);
      \draw[S] (11) -- (19);
      \draw[S] (6) -- (11);
      \draw[S] (4) -- (7);
      \draw[S] (7) -- (13);
      \draw[S] (7) -- (9);
      \draw[S] (13) -- (15);
      \draw[S] (13) -- (18);
      \draw[S] (15) -- (17);
      \draw[S] (15) -- (21);
      \draw[S] (15) -- (22);
      \draw[S] (15) -- (20);

      \draw[G] (0) to (1);
      \draw[G] (1) to (11);
      \draw[G] (6) to (11);
      \draw[G] (19) to (11);
      \draw[G] (2) to (6);
      \draw[G] (2) to (8);
      \draw[G] (8) to (12);
      \draw[G] (8) to (16);
      \draw[G] (2) to (10);
      \draw[G] (3) to (10);
      \draw[G] (3) to (14);
      \draw[G] (1) to (21);
      \draw[G] (21) to (15);
      \draw[G] (15) to (22);
      \draw[G] (15) to (20);
      \draw[G,bend left=7] (15) to (17);
      \draw[G] (15) to (13);
      \draw[G] (15) to (7);
      \draw[G] (15) to (4);
      \draw[G] (4) to (5);
      \draw[G] (7) to (9);
      \draw[G] (13) to (18);
      \draw[G] (10) to (14);
      \draw[G,bend left=20] (0) to (21);

      \draw[U] (0) to (1);
      \draw[U] (1) to (11);
      \draw[U] (6) to (11);
      \draw[U] (19) to (11);
      \draw[U] (2) to (6);
      \draw[U] (2) to (8);
      \draw[U] (8) to (12);
      \draw[U] (8) to (16);
      \draw[U] (2) to (10);
      \draw[U] (3) to (10);
      \draw[U] (10) to (14);
      \draw[U,bend left=15] (1) to (21);
      \draw[U] (21) to (15);
      \draw[U] (15) to (22);
      \draw[U] (15) to (20);
      \draw[U,bend right=7] (15) to (17);
      \draw[U,bend left=15] (15) to (4);
      \draw[U,bend left=15] (15) to (7);
      \draw[U] (15) to (13);
      \draw[U] (13) to (18);
      \draw[U] (7) to (9);
      \draw[U] (4) to (5);
    \end{tikzpicture}
    \caption[A graph $G$, the elimination tree $S$ and a tree $U$]%
    {A graph $G$ (solid black lines), the elimination tree $S$ (dashed blue
      lines), and the tree $U$ (dotted red lines). Numbering of nodes
      reflects the ordering $L$.}
    \label{fig:G_S_U}
  \end{figure*}

  Let $S=S(G,L)$ be the rooted elimination tree of $G$ imposed by $L$. For
  a vertex $u$, by $G_u$ we denote the subgraph of $G$ induced by all
  descendants of $u$ in $S$, including $u$. The following properties follow
  easily from the construction of a rooted elimination tree.

  \begin{claim}\label{prop:eltree} \
    The following assertions hold.

    \smallskip\qitem{1.}{For each $u\in V(G)$, the subgraph $G_u$ is
      connected.}

    \smallskip\qitem{2.}{Whenever a vertex $u$ is an ancestor of a vertex
      $v$ in $S$, we have $u\le_Lv$.}

    \smallskip\qitem{3.}{For each $uv\in E(G)$ with $u<_Lv$, $u$ is an
      ancestor of~$v$ in $S$.}

    \smallskip\qitem{4.}{For each $u\in V(G)$ and each child $v$ of $u$ in
      $S$, $u$ has at least one neighbour in $V(G_v)$.}
  \end{claim}

  \begin{clproof}\quad Assertions~1 and~2 follow immediately from the
    construction of $S$. For assertion~3, suppose that $u$ and $v$ are not
    bound by the ancestor-descendant relation in $S$, and let $w$ be their
    lowest common ancestor in $S$. Then $u$ and $v$ would be in different
    connected components of $G_w-w$, hence $uv$ could not be an edge; a
    contradiction. It follows that $u$ and $v$ are bound by the
    ancestor-descendant relation, implying that~$u$ is an ancestor of $v$,
    due to $u<_L v$ and assertion~2. Finally, for assertion~4, recall that
    by assertion~1 we have that $G_u$ is connected, whereas by construction
    $G_v$ is one of the connected components of $G_u-u$. Hence, in $G$
    there is no edge between $V(G_v)$ and any of the other connected
    components of $G_u-u$. If there was no edge between $V(G_v)$ and $u$ as
    well, then there would be no edge between $V(G_v)$ and $V(G_u)\setminus
    V(G_v)$, contradicting the connectivity of $G_u$.
  \end{clproof}

  \noindent
  We now define a set of edges $B\subseteq E(G)$ as follows. For every
  vertex $u$ of $G$ and every child $v$ of~$u$ in $S$, select an arbitrary
  neighbour $w_{u,v}$ of $u$ in $G_v$; such a neighbour exists by
  \cref{prop:eltree}.4. Then let $B_u$ be the set of all edges $uw_{u,v}$,
  for~$v$ ranging over the children of $u$ in~$S$. Define
  \[B=\bigcup_{u\in V(G)} B_u.\]
  Let $U$ be the graph spanned by all the edges in $B$, that is,
  $U=(V(G),B)$. In \Cref{fig:G_S_U}, the edges of $U$ are represented by
  the dotted red lines.

  \begin{claim}\label{lem:is-tree} \
    The graph $U$ is a tree.
  \end{claim}

  \begin{clproof}\quad Observe that for each $u\in V(G)$, the number of
    edges in $B_u$ is equal to the number of children of $u$ in $S$. Since
    every vertex of $G$ has exactly one parent in $S$, apart from the root
    of $S$, we infer that
    \[|B|\le \sum_{u\in V(G)}|B_u|= |V(G)|-1.\]
    Therefore, since $B$ is the edge set of $U$, to prove that~$U$ is a
    tree it suffices to prove that $U$ is connected. To this end, we prove
    by a bottom-up induction on $S$ that for each $u\in V(G)$, the subgraph
    $U_u=\bigl(V(G_u),B\cap \binom{V(G_u)}{2}\bigr)$ is connected. Note
    that for the root~$w$ of~$S$ this claim is equivalent to $U_w=U$ being
    connected.

    Take any $u\in V(G)$, and suppose by induction that for each child $v$
    of~$u$ in $S$, the subgraph~$U_v$ is connected. Observe that~$U_u$ can
    be constructed by taking the vertex $u$ and, for each child $v$ of~$u$
    in $S$, adding the connected subgraph $U_v$ and connecting it to $u$
    via edge $uw_{u,v}\in B_u$. Thus, $U_u$ constructed in this manner is
    also connected, as claimed.
  \end{clproof}

  \noindent
  Next, we verify that $U$ can be computed within the claimed running time.
  Note that we do not need to compute $S$, as we will use it only in the
  analysis. We remark that this is the only place in the algorithm where
  the running time is not linear.

  \begin{claim} \
    The tree $U$ can be computed in time $\Oh(m\cdot\alpha(m))$.
  \end{claim}

  \begin{clproof}\quad We use the classic Find\,\&\,Union data structure on
    the set $V(G)$. Recall that in this data structure, at each moment we
    maintain a partition of $V(G)$ into a number of equivalence classes,
    each with a prescribed representative, where initially each vertex is
    in its own class. The operations are a) for a given $u\in V(G)$, find a
    representative of the class to which $u$ belongs, and b) merge two
    equivalence classes into one. Tarjan~\cite{Tarjan75} gave an
    implementation of this data structure where both operations run in
    amortised time $\alpha(k)$, where $k$ is the total number of operations
    performed.

    Having initialised the data structure, we process the vertex ordering
    $L$ from the smallest end, starting with an empty suffix. For an
    already processed suffix $X$ of $L$, the maintained classes within $X$
    will represent the partition of $G[X]$ into connected components, while
    every vertex outside~$X$ will still be in its own equivalence class.
    Let us consider one step, when we process a vertex $u$, thus moving
    from a suffix $X$ to the suffix $X'=X\cup \{u\}$. Iterate through all
    the neighbours of~$u$, and for each neighbour $v$ of $u$ such that
    $u<_L v$, verify whether the equivalence classes of $u$ and $v$ are
    different. If this is the case, merge these classes and add the edge
    $uv$ to $B$. A straightforward induction shows that the claimed
    invariant holds. Moreover, when processing $u$ we add exactly the edges
    of $B_u$ to $B$, hence at the end we obtain the set $B$ and the tree
    $U=(V(G),B)$.

    For the running time analysis, observe that in total we perform
    $\Oh(m)$ operations on the data structure, thus the running time is
    $\Oh(m\alpha(m))$. We remark that we assume that the ordering $L$ is
    given as a bijection between $V(G)$ and numbers
    $\{1,2,\ldots,|V(G)|\}$, thus for two vertices $u,v$ we can check in
    constant time whether $u<_L v$.
  \end{clproof}
  
  \noindent
  By \cref{lem:is-tree} we have that $U$ is a spanning tree of~$G$, however
  its maximum degree may be (too) large. The idea is to use $U$ to
  construct a new tree~$T$ with maximum degree at most~$3$ (on the same
  vertex set $V(G)$). The way we constructed~$U$ will enable us to argue
  that adding the edges of~$T$ to the graph $G$ does not change the
  generalised colouring numbers too much.

  Give $U$ the same root as the elimination tree $S$. From now on we treat
  $U$ as a rooted tree, which imposes parent-child and
  an\-ces\-tor-de\-s\-cen\-dant relations in $U$ as well. Note that the
  parent-child and an\-ces\-tor-de\-s\-cen\-dant relations in~$S$ and in
  $U$ may be completely different. For instance, consider vertices $4$ and
  $15$ in the example from \Cref{fig:G_S_U}: $4$ is a child of $15$ in $U$,
  and an ancestor of $15$ in $S$.

  \begin{figure*}
    \centering
    \begin{tikzpicture}[scale=1.1]
      \node[vertex] (0) at (0,0){$0$};
      \node[vertex] (1) at (0,-1){$1$};
      \node[vertex] (2) at (-3,-2){$2$};
      \node[vertex] (3) at (-6,-3){$3$};
      \node[vertex] (4) at (4,-2){$4$};
      \node[vertex] (5) at (7,-4.5){$5$};
      \node[vertex] (6) at (-1,-3){$6$};
      \node[vertex] (7) at (4,-3){$7$};
      \node[vertex] (8) at (-3,-3){$8$};
      \node[vertex] (9) at (6,-4.5){$9$};
      \node[vertex] (14) at (-6,-5){$14$};
      \node[vertex] (19) at (-1,-5){$19$};
      \node[vertex] (12) at (-4,-4){$12$};
      \node[vertex] (13) at (4,-4){$13$};
      \node[vertex] (10) at (-5,-4){$10$};
      \node[vertex] (15) at (3,-5){$15$};
      \node[vertex] (16) at (-2,-4){$16$};
      \node[vertex] (17) at (5.5,-6){$17$};
      \node[vertex] (18) at (5,-4.5){$18$};
      \node[vertex] (11) at (0,-4){$11$};
      \node[vertex] (20) at (4,-6){$20$};
      \node[vertex] (21) at (2,-6){$21$};
      \node[vertex] (22) at (3,-6){$22$};

      \draw[G] (0) to (1);
      \draw[G] (1) to (11);
      \draw[G] (6) to (11);
      \draw[G] (19) to (11);
      \draw[G] (2) to (6);
      \draw[G] (2) to (8);
      \draw[G] (8) to (12);
      \draw[G] (8) to (16);
      \draw[G] (2) to (10);
      \draw[G] (3) to (10);
      \draw[G] (3) to (14);
      \draw[G] (1) to (21);
      \draw[G] (21) to (15);
      \draw[G] (15) to (22);
      \draw[G] (15) to (20);
      \draw[G,bend left=7] (15) to (17);
      \draw[G] (15) to (13);
      \draw[G] (15) to (7);
      \draw[G,bend left=30] (15) to (4);
      \draw[G] (4) to (5);
      \draw[G] (7) to (9);
      \draw[G] (13) to (18);
      \draw[G] (10) to (14);
      \draw[G,bend left=20] (0) to (21);

      \draw[U] (0) to (1);
      \draw[U] (1) to (11);
      \draw[U] (6) to (11);
      \draw[U] (19) to (11);
      \draw[U] (2) to (6);
      \draw[U] (2) to (8);
      \draw[U] (8) to (12);
      \draw[U] (8) to (16);
      \draw[U] (2) to (10);
      \draw[U] (3) to (10);
      \draw[U] (10) to (14);
      \draw[U,bend left=15] (1) to (21);
      \draw[U] (21) to (15);
      \draw[U] (15) to (22);
      \draw[U] (15) to (20);
      \draw[U,bend right=7] (15) to (17);
      \draw[U,bend left=15] (15) to (4);
      \draw[U,bend left=15] (15) to (7);
      \draw[U] (15) to (13);
      \draw[U] (13) to (18);
      \draw[U] (7) to (9);
      \draw[U] (4) to (5);

      \draw[T] (0) to (1);
      \draw[T] (1) to (11);
      \draw[T] (6) to (11);
      \draw[T] (6) to (19);
      \draw[T] (2) to (6);
      \draw[T] (10) to (8);
      \draw[T] (8) to (12);
      \draw[T] (12) to (16);
      \draw[T] (2) to (8);
      \draw[T] (3) to (10);
      \draw[T] (3) to (14);
      \draw[T] (11) to (21);
      \draw[T] (21) to (15);
      \draw[T] (13) to (17);
      \draw[T] (22) to (20);
      \draw[T] (20) to (17);
      \draw[T,bend left=23] (15) to (4);
      \draw[T] (4) to (7);
      \draw[T] (7) to (13);
      \draw[T] (13) to (18);
      \draw[T] (7) to (9);
      \draw[T] (4) to (5);
    \end{tikzpicture}
    \caption[The graph $G$, the tree $U$ and a tree $T$]%
    {A graph $G$ (solid black lines), the tree $U$ (dotted red lines), and
      the tree $T$ (thick dashed green lines).}
    \label{fig:G_U_T}
  \end{figure*}
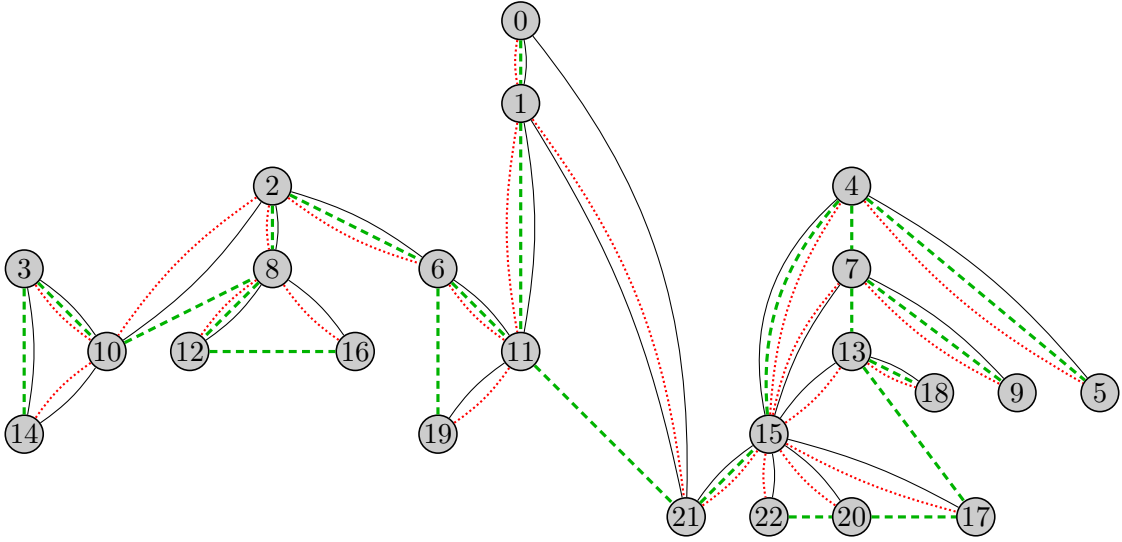

  For every $u\in V(G)$, let $(x_1,\ldots,x_p)$ be an enumeration of the
  children of $u$ in $U$, such that $x_i<_L x_j$ if $i<j$. Let
  $F_u=\{ux_1,x_1x_2,x_2x_3,\ldots,x_{p-1}x_{p}\}$, and define
  \[F= \bigcup_{u\in V(G)}F_u\quad \text{and}\quad T=(V(G),F).\]
  See \Cref{fig:G_U_T} for an illustration.

  \begin{claim}\label{lem:correctness} \
    The graph $T$ is a tree with maximum degree at most~$3$.
  \end{claim}

  \begin{clproof}\quad Observe that for each $u\in V(G)$, we have that
    $|F_u|$ is equal to the number of children of~$u$ in $U$. Every vertex
    of $G$ apart from the root of $U$ has exactly one parent in $U$, hence
    \[|F|\le \sum_{u\in V(G)}|F_u|=V(G)|-1.\]
    Therefore, to prove that $T$ is a tree, it suffices to argue that it is
    connected. This, however, follows immediately from the fact that $U$ is
    connected, since for each edge in $U$ there is a path in $T$ that
    connects the same pair of vertices.

    Finally, it is easy to see that each vertex $u$ is incident to at most
    $3$ edges of $F$: at most one leading to a child of $u$ in~$U$, and at
    most $2$ belonging to $F_v$, where $v$ is the parent of $u$ in~$U$.
  \end{clproof}
  
  \noindent
  Observe that once the tree $U$ is constructed, it is straightforward to
  construct $T$ in time $\Oh(n)$. Thus, it remains to check that adding $F$
  to $G$ does not change the generalised colouring numbers too much.

  Take any vertex $u\in V(G)$ and examine its children in~$U$. We partition
  them as follows. Let~$\Zup_u$ be the set of those children of~$u$ in $U$
  that are its ancestors in $S$, and let $\Zdown_u$ be the set of those
  children of $u$ in $U$ that are its descendants in~$S$. By the
  construction of $U$ and by \cref{prop:eltree}.3, each child of~$u$ in $U$
  is either its ancestor or descendant in $S$. By \cref{prop:eltree}.2,
  this is equivalent to saying that $\Zup_u$, respectively $\Zdown_u$,
  comprise the children of $u$ in $U$ that are smaller, respectively
  larger, than $u$ in $L$. Note that by the construction of $U$, the
  vertices of $\Zdown_u$ lie in pairwise different subtrees rooted at the
  children of $u$ in $S$, thus $u$ is the lowest common ancestor in $S$ of
  every pair of vertices from $\Zdown_u$. On the other hand, all vertices
  of $\Zup_u$ are ancestors of $u$ in~$S$, thus every pair of them is bound
  by the an\-ces\-tor-de\-s\-cen\-dant relation in $S$.

  \begin{claim}\label{lem:adm-bound} \
    The graph $G+F$ satisfies the following inequality:
    $\adm_r(G+F,L)\le 3\cdot\col_{2r}(G,L)$.
  \end{claim}

  \begin{clproof}\quad Write $H=G+F$. Let $\Fnew=F\setminus E(G)$ be the
    set of edges from $F$ that were not already present in $G$. If an edge
    $e\in\Fnew$ belongs also to $F_u$ for some $u\in V(G)$, then we know
    that $u$ cannot be an endpoint of $e$. This is because each edge
    joining a vertex $u$ with one of its children in~$U$ is already present
    in $G$. We say that the vertex~$u$ is the \emph{origin} of an edge
    $e{}\in\Fnew\cap F_u$, and denote it by $a(e)$. Observe that then both
    endpoints of $e$ are the children of~$a(e)$ in the tree~$U$, and hence
    $a(e)$ is adjacent to both the endpoints of $e$ in $G$.

    To give an upper bound on $\adm_r(H,L)$, let us fix a vertex $u\in
    V(G)$ and a family of paths~$\Pp$ in $H$ such that

    \smallskip\qitem{$\bullet$}{each path in $\Pp$ has length at most $r$,
      starts in $u$, ends in a vertex smaller than $u$ in $L$, and all its
      internal vertices are larger than $u$ in $L$;}

    \smallskip\qitem{$\bullet$}{the paths in $\Pp$ are pairwise
      vertex-disjoint, apart from the starting vertex $u$.}

    \smallskip\noindent
    For each path $P\in \Pp$, we define a walk $P'$ in $G$ as follows. For
    every edge $e=xy$ from~$\Fnew$ traversed on $P$, replace the usage of
    this edge on $P$ by the following detour of length $2$: $x{-}a(e){-}y$.
    Observe that $P'$ is a walk in the graph $G$, it starts in $u$, ends in
    the same vertex as $P$, and has length at most~$2r$. Next, we define
    $v(P)$ to be the first vertex on $P'$ (that is, the closest to~$u$ on
    $P'$) that does not belong to $G_u$. Since the endpoint of $P'$ that is
    not $u$ does not belong to~$G_u$, such a vertex exists. Finally, let
    $P''$ be the prefix of $P'$ from $u$ to the first visit of $v(P)$ on
    $P'$ (from the side of $u$). Observe that the predecessor of $v(P)$ on
    $P''$ belongs to $G_u$ and is a neighbour of $v(P)$ in $G$, hence
    $v(P)$ has to be a strict ancestor of $u$ in~$S$. We find that $P''$ is
    a walk of length at most $2r$ in $G$, it starts in $u$, ends in $v(P)$,
    and all its internal vertices belong to $G_u$, so in particular they
    are not smaller than $u$ in $L$. This means that~$P''$ certifies that
    $v(P)\in\Sreach_{2r}[G,L,u]$.

    Since $|\Sreach_{2r}[G,L,u]|\le \col_{2r}(G,L)$, in order to prove the
    bound on $\adm_r(H,L)$, it suffices to prove the following claim: For
    each vertex $v$ that is a strict ancestor of $u$ in~$S$, there can be
    at most three paths $P\in\Pp$ for which $v=v(P)$. To this end, we fix a
    vertex $v$ that is a strict ancestor of~$u$ in~$S$ and proceed by a
    case distinction on how a path $P$ with $v=v(P)$ may behave.

    Suppose first that $v$ is the endpoint of $P$ other than $u$,
    equivalently the endpoint of $P'$ other than $u$. (For example, $u=1$,
    $P = 1,11,21,0$, $P' = 1,11,1,21,0$ and $v=0$, in
    \Cref{fig:G_S_U,fig:G_U_T}.) However, the paths of $\Pp$ are pairwise
    vertex-disjoint, apart from the starting vertex $u$, hence there can be
    at most one path $P$ from $\Pp$ for which $v$ is an endpoint. Thus,
    this case contributes at most one path $P$ for which $v=v(P)$.

    Next suppose that $v$ is an internal vertex of the walk $P'$; in
    particular, it is not the endpoint of~$P$ other than $u$. (For example,
    $u=6$, $P = 6,11,21,0$, $P' = 6,11,1,21,0$ and $v = 1$, in
    \Cref{fig:G_S_U,fig:G_U_T}.) Since the only vertex traversed by~$P$
    that is smaller than $u$ in $L$ is this other endpoint of $P$, and $v$
    is smaller than $u$ in $L$ due to being its strict ancestor in $S$, it
    follows that each visit of $v$ on $P'$ is due to having $v=a(e)$ for
    some edge $e\in\Fnew$ traversed on~$P$. Select $e$ to be such an edge
    corresponding to the first visit of $v$ on $P'$. Let $e=xy$, where $x$
    lies closer to $u$ on $P$ than $y$. (That is, in our figures, $x=11$
    and $y=21$.) Since $v$ was chosen as the first vertex on $P'$ that does
    not belong to $G_u$, we have $x\in G_u$.

    Since $v=a(e)=a(xy)$, either $x\in\Zdown_v$ or $x\in\Zup_v$. Note that
    the second possibility cannot happen, because then~$v$ would be a
    descendant of $x$ in $S$, hence $v$ would belong to~$G_u$, due to
    $x\in G_u$; a contradiction. This means $x\in\Zdown_v$.

    Recall that, by construction, $\Zdown_v$ contains at most one vertex
    from each subtree of $S$ rooted at a child of $v$. Since $v$ is a
    strict ancestor of $u$ in $S$, we infer that $x$ has to be the unique
    vertex of $\Zdown_v$ that belongs to $G_u$. In the construction of
    $F_v$, however, we added only at most two edges of $F_v$ incident to
    this unique vertex: at most one to its predecessor on the enumeration
    of the children of $v$, and at most one to its successor. Since paths
    from $\Pp$ are pairwise vertex-disjoint in~$H$, apart from the starting
    vertex $u$, only at most two paths from $\Pp$ can use any of these two
    edges (actually, only at most one unless $x = u$). Only for these two
    paths we can have $v=a(e)$. Thus, this case contributes at most two
    paths $P$ for which $v=v(P)$, completing the proof of the claim.
  \end{clproof}

  \noindent
  We conclude the proof by summarising the algorithm: first construct the
  tree $U$, and then construct the tree $T$. As argued, these steps take
  time $\Oh(m\cdot\alpha(m))$ and $\Oh(n)$, respectively. By
  Claims~\ref{lem:correctness} and~\ref{lem:adm-bound}, $T$ satisfies the
  required properties.
\end{proof}

\section{Conclusion}

In this paper we show that model-checking for successor-invariant
first-order formulas is fixed-parameter tractable on any class of
structures of bounded expansion. This significantly reduces the existing
gap for sparse classes between the known tractability results for plain and
for successor-invariant first-order logic.

The obvious open question is whether this gap can be closed completely on
sparse classes of graphs, i.e.\ whether successor-invariant \FO\ is
fixed-parameter tractable on any nowhere dense class of structures. As
nowhere dense classes can also be characterised by colouring numbers, it is
conceivable that our techniques can be extended. However, for nowhere dense
classes of graphs, the colouring numbers are no longer bounded by a
constant (for any fixed value of $r$) but only by $n^\epsilon$, where $n$
is the number of vertices. This poses several technical problems meaning
that our techniques do not readily extend. We leave this as an open problem
for future research.

Another open problem is the exact time complexity of our model-checking
algorithm on classes of bounded expansion. For plain \FO\ it is known that
$\MC(\FO,\CCC)$ is parametrised linear time for any class $\CCC$ of
bounded expansion.

The simple analysis of our algorithm provided in this paper yields a
parametrised running time of $\Oh(n\cdot\alpha(n))$. The only step that
requires more than linear time is the construction of a specific spanning
tree in \Cref{thm:main-technical}. At the moment we do not see how to avoid
this non-linear step, and we leave it for future research whether the tree
$T$ in the theorem (or a similar suitable tree) can be constructed more
efficiently.

Finally, it would be very interesting to extend our results to
order-invariant \FO. One approach would be to show that model-checking for
\FO\ augmented by a reachability operator that can define reachability
along the (definable) successor-relation constructed above is tractable on
bounded expansion classes. However, as explained in the introduction,
first-order logic with even very restricted forms of reachability very
quickly becomes intractable even on planar graphs. Again, this is an area
where further research seems appropriate.

\bibliographystyle{plain}
\bibliography{ref}

\end{document}